\documentclass[prb,aps,twocolumn,floats,showpacs]{revtex4-2}
\usepackage{amsthm}     
\usepackage{amsmath,amssymb,bm}
\usepackage{graphicx}
\usepackage{caption}
\usepackage{float}
\usepackage{makecell}
\usepackage{pifont}
 \usepackage{ragged2e}
 \usepackage{booktabs}   
\usepackage{multirow}   
\usepackage{xcolor}     
\usepackage[colorlinks=true, letterpaper=true, pdfstartview=FitV, linkcolor=blue, citecolor=blue, urlcolor=blue]{hyperref}

\usepackage{color}

\def\eqa{\begin{eqnarray}}
\def\eea{\end{eqnarray}}
\newcommand{\eq}{\begin{equation}}
\newcommand{\ee}{\end{equation}}

\usepackage{physics}

\begin{document}
\title{Chiral phonon induced electron anomalous Hall effect in monolayer transition metal Dichalcogenides}
\author{Ka-San Mak$^{1,2}$}
\author{Jia-Xing Zhang$^{3}$}
\author{Wei Chen$^{1,4}$} \email{chenweiphy@nju.edu.cn}
\affiliation{$^{1}$National Laboratory of Solid State Microstructures and School of Physics, Nanjing University, Nanjing, China}
\affiliation{$^{2}$Department of Physics, Faculty of Science, University of Macau, Macau, China}
\affiliation{$^{3}$School of Physics and Astronomy, Shanghai Jiaotong University, Shanghai, China}
\affiliation{$^{4}$Collaborative Innovation Center of Advanced Microstructures, Nanjing University, Nanjing, China}

\begin{abstract}
We study the chiral phonon induced electron anomalous Hall effect in monolayer transition metal Dichalcogenides with hole doping in this work. The chiral phonon has been observed in this system in experiment previously. The electron Hamiltonian of the system breaks inversion symmetry but preserves time reversal symmetry. The strong spin-orbit coupling in the valence band results in spin polarization in each valley and spin-valley locking. We show that though the intrinsic electron anomalous Hall effect vanishes in this system due to the time reversal symmetry of the electron Hamiltonian, the extrinsic contribution to the anomalous Hall conductivity due to chiral phonon scatterings with the electrons is finite because the chiral phonon breaks time reversal symmetry in the system. We studied both the defomation potential and vector potential couplings of the chiral phonon and electrons and showed that the anomalous charge Hall effect due to deformation potential coupling in the two valleys cancel each other in the system. However, the total anomalous charge Hall effect of the two valleys due to vector potential coupling  is finite. Besides, the inversion symmetry breaking in the system also results in significant valley Hall effect which includes both intrinsic and extrinsic contribution and in turn leads to spin Hall effect due to the spin-valley locking. 
\end{abstract}

\maketitle  

\newpage

\section{Introduction}

The physics of chiral phonons has emerged as a rapidly developing frontier in condensed matter physics~\cite{Sheng2006, Mori2014, Zhang2010, ZhangNiu2014, ZhangNiu2015, Zhu2018, Ren2021, Dominik2025, Zhang2025, Choi2024,Kim2026,Sun2026, Chen2024}. In contrast to conventional linearly polarized phonons, chiral phonons carry angular momentum and chirality and can break time-reversal symmetry (TRS) by lifting the degeneracy between left- and right-handed modes\cite{Sheng2006, Mori2014, ZhangNiu2014,ZhangNiu2015}. The unique properties of chiral phonons give rise to a variety of intriguing phenomena, including the phonon Hall effect \cite{Strohm2005, Sheng2006, Zhang2010, Mori2014}, phonon magnetic moments\cite{Spaldin2019, Grissonnanche2020, Ren2021, Wang2024, Lin2025, Wang2025} {\it etc}.. Furthermore, the coupling between chiral phonons and electrons can have profound consequences on the electronic properties, leading to phenomena such as chiral phonon-induced electron spin magnetization\cite{Fransson2023, Yao2025}, chiral-phonon-driven currents in helical systems\cite{YaoMurakami2022}.

In this work, we present another distinctive consequence of chiral phonon-electron interaction in monolayer transition metal dichalcogenides (TMDs): the electron anomalous Hall effect (AHE) with time reversal invariant electron Hamiltonian. 
The monolayer TMD has attracted a lot of interest in recent years due to its fascinating phononic and electronic properties~\cite{Zhu2018, Novoselov2005, Wang2010, Mak2010, Xiao2012, Guinea2014, Shan2020, Shan2022}. 
The existence of chiral phonons was first observed experimentally in this sytem in 2018~\cite{Zhu2018}. 
On the electron side, the strong spin-orbit coupling in the valence band of this system polarizes the spin in the upmost valence band in each valley and results in spin-valley locking~\cite{Zhu2011, Xiao2012}. 
The electron Hamiltonian of monolayer TMDs breaks inversion symmetry but preserves TRS~\cite{Xiao2012, Guinea2014}. It is well established that the electron AHE generally requires TRS breaking and includes intrinsic and extrinsic contributions~\cite{Sinitsyn2007, Sinitsyn2008}. The former originates from the non-trivial electronic band structure, while the latter arises from electron scatterings by external degrees of freedom, such as impurities. In previous studies of the electron AHE, TRS breaking has typically been incorporated directly into the electron Hamiltonian~\cite{Sinitsyn2007, Chen2023, Titov2016, ZhangChen2024}. Here, we demonstrate that an AHE can nevertheless arise even when the electron Hamiltonian itself preserves TRS. Specifically, although the preservation of TRS in monolayer TMDs ensures that the intrinsic contribution to the electron AHE vanishes, the extrinsic contribution arising from electron scatterings by chiral phonons remains finite because the chiral phonons break TRS of the system.

We study two types of chiral phonon-electron couplings in the monolayer TMD: the deformation potential (DP) and pseudo-gauge vector potential (VP) couplings~\cite{Guinea2014, Shan2020, Shan2022}. 
We show that for Fermi energy locating at the upmost valence band of monolayer TMD, the pure DP scattering induced electron anomalous Hall conductivity has the same value but opposite sign in the two valleys so the total charge Hall conductivity from pure DP coupling is zero. However, the VP induced anomalous charge Hall conductivity has both different values and signs in the two valleys, therefore the total charge Hall conductivity is finite. Other than the anomalous charge Hall effect, the chiral phonon-electron coupling also results in significant valley Hall effect in monolayer TMD due to the inversion symmetry breaking in the system~\cite{Xiao2007}, for which both the DP and VP scatterings result in finite contribution. The valley Hall effect also accompanies a significant spin Hall effect in the system due to the spin-valley locking in the system. In contrast to the anomalous charge Hall effect, the valley and spin Hall effect includes both  intrinsic and extrinsic contributions in monolayer TMD.

This paper is organized as follows. In Sec.\ref{Model}, we present the electron and chiral phonon model in monolayer TMD as well as the DP and VP coupling between them.
In Sec.\ref{AHE}, we present the calculation of the chiral phonon induced electron anomalous Hall effect in monolayer TMD, as well as the valley and spin Hall effect in this system. In Sec.\ref{Discussion}, we have a discussion of some relevant issues and 
conclude the work.

\begin{figure}[t] 
   \includegraphics[width=0.45\textwidth]{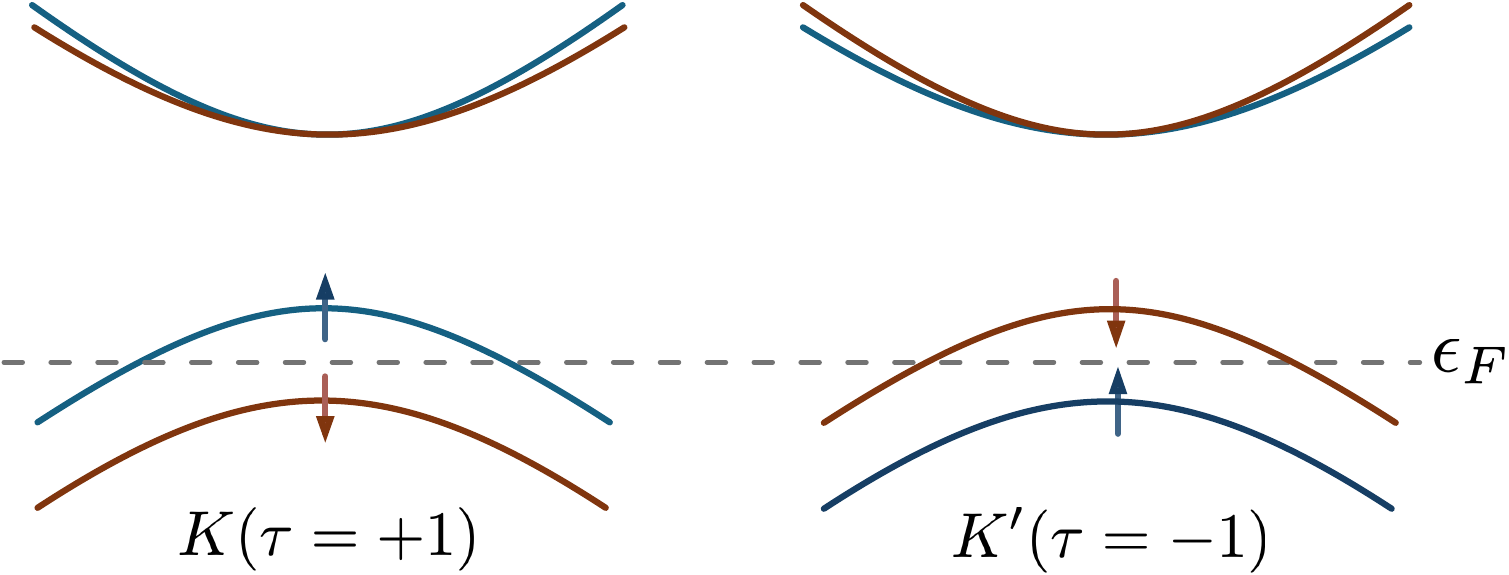} 
 \caption{  \justifying Schematic band structure of monolayer transition metal Dichalcogenides. The valence band splits to two bands with spin up and spin down due to the spin-orbit coupling. The spin splitting is opposite in the two valleys. For Fermi energy $\epsilon_F$ locating at the topmost valence band in the figure, the spin is polarized on the Fermi surface in each valley with opposite directions in the two valleys due to time reversal symmetry. }
  \label{fig:band}
\end{figure}

\section{Model and Hamiltonians}\label{Model}
The electron band structure of monolayer TMD has been studied in \cite{Xiao2012}. The inversion symmetry is broken in the system and the $d$ orbitals of the heavy metal atoms lead to strong spin-orbit coupling in the valence band~\cite{Zhu2011}.
The conduction and valence band of the monolayer TMDs near the band edge $K$ points of the hexagonal Brillouin zone  can be described by a massive Dirac model with spin-orbit interaction at each valley as~\cite{Xiao2012}
\begin{equation}\label{H_soc}
\hat{H}(k)=v(\tau k_x\hat{\sigma}_x+k_y\hat{\sigma}_y)+\frac{\Delta}{2}\hat{\sigma}_z
-\frac{\lambda_{\mathrm{so}}\tau \hat{s}_z}{2}(\hat{\sigma}_z-I),
\end{equation}
where $\tau=\pm 1$ labels the two valleys,  
$v$ is electron velocity, $k_x, k_y$ are the electron momenta,
$\Delta$ the energy gap and 
 $\hat{\sigma}_{x, y, z}$ denote the Pauli matrices acting on the spinor basis $|\psi_\tau\rangle=(|\psi_{\tau, c} \rangle, |\psi_{\tau,v}\rangle)^T$ with $|\psi_{\tau,c} \rangle$ and $|\psi_{\tau, v}\rangle$ the conduction and valence band electron states at valley $\tau$ respectively.
  The last term in Eq.(\ref{H_soc}) describes the spin-orbit coupling, with $\lambda_{\mathrm{so}}$ the coupling constant, $\hat{s}_z$ the $z$-component spin operator with eigenvalue $s_z=\pm \frac12$. The spin-orbit coupling in the conduction band is weak so is ignored in the Hamiltonian. 
  Under time reversal,  $\tau, \hat{s}_z$ and $(k_x, k_y)$ changes sign, and  $(\hat{\sigma}_x, \hat{\sigma}_y, \hat{\sigma}_z)$
 transforms to $(\hat{\sigma}_x, -\hat{\sigma}_y, \hat{\sigma}_z)$. 
 Under inversion operation, $(\hat{\sigma}_x, \hat{\sigma}_y, \hat{\sigma}_z)$ and $\hat{s}_z$ remains the same whereas $(k_x, k_y)$ and $\tau$ changes sign.
 The Hamiltonian then preserves TRS but breaks inversion symmetry. 
 

The spectrum of Hamiltonian Eq.(\ref{H_soc}) is 
\begin{equation}
\epsilon_{\tau, s,\eta}(\mathbf{k})=\varepsilon_{\tau s}+\eta E_{\tau s}(\mathbf{k}),\ E_{\tau s}(\mathbf{k})\equiv\sqrt{v^2k^2+ m_{\tau s}^2},
\end{equation}
with $\eta=\pm 1$ labeling the two bands, $s=\pm 1$ labeling the spin-up and spin-down components, $\varepsilon_{\tau s}=\frac{\lambda_{\rm so}\tau s_z}{2} $ and $m_{\tau s}=\frac{\Delta-\lambda_{\rm so}\tau s_z}{2}$. 
The spin-up and spin-down components in the original valence band then  split to two bands and are decoupled as shown in Fig.\ref{fig:band}. 
For Fermi energy $\epsilon_F$ locating at the topmost valence band shown in Fig.\ref{fig:band}, the spin is polorized on the Fermi surface with opposite polarization in the two valleys, i.e., the spin and valley is locked.

The chiral phonons in monolayer TMDs have been theoretically studied in \cite{ZhangNiu2015} and observed in experiments \cite{Zhu2018}. There are both acoustic and optical chiral phonon modes in this system\cite{ZhangNiu2015}. For simplicity, in this work we focus on the study of the acoustic chiral phonon interaction with the electrons. The chiral phonons can originate from an effective vector potential $\mathbf{A}(\mathbf{R}_i)$ the electrons produced due to ionic oscillations~\cite{Niu1998, Vanderbilt2024}. An effective Hamiltonian of the acoustic chiral phonon mode can be written as~\cite{Vanderbilt2024} 
\begin{eqnarray}\label{H_phonon}
\hat{H}_{\rm ph} &=
\sum_{i}\frac{1}{2}\left[(\hat{p}^x_i)^2+(\hat{p}^y_i)^2\right]
+\sum_{i}\frac{1}{2}\left[\omega_{x}^2(\hat{u}^x_i)^2+
\omega_{y}^2(\hat{u}^y_i)^2\right] \nonumber\\
&-\sum_{i}\frac{\lambda_{i}}{2}\hat{L}^z_i,
\end{eqnarray}
where $i$ labels the lattice sites, $\hat{p}_i^{\alpha}, \hat{u}_i^{\alpha}, \alpha=x, y$ are the momentum and displacement of the ion on site $i$ in the $x$ and $y$ directions respectively, and 
\begin{equation}
\hat{L}^z_i=\hat{u}^x_i \hat{p}^y_i-\hat{u}^y_i \hat{p}^x_i
\end{equation}
is the $z$ component of the angular momentum operator of the ion. It couples to the $z$ component of an effective Berry curvature  $\lambda_{i}=\Omega^z(\mathbf{R}_i)$~\cite{Niu1998} with 
\begin{equation}
\mathbf{\Omega}(\mathbf{R}_i)=\triangledown\cross \mathbf{A}(\mathbf{R}_i),
\ \ \mathbf{A}(\mathbf{R}_i)= i\langle \Psi_e(\mathbf{R}_i)|\frac{\partial \Psi_e(\mathbf{R}_i)}{\partial \mathbf{u}_i}\rangle
\end{equation}
where $\mathbf{R}_i=\mathbf{R}^0_i+\mathbf{u}_i$ is the coordinate of the ion on site $i$ and $\mathbf{u}_i$ is the displacement. 

The Hamiltonian Eq.(\ref{H_phonon}) after Fourier transformation  becomes
\begin{eqnarray}\label{H_k_phonon}
\hat{H}_{\rm ph}&=&\frac{1}{2}\sum_q \left[\hat{\mathbf p}^T_{-\mathbf{q}}\cdot\hat{\mathbf p}_\mathbf{q}+\omega_\mathbf{q}^2 \hat{\mathbf u}^T_{-\mathbf{q}}\cdot\hat{\mathbf u}_\mathbf{q}\right] \nonumber\\
&&
\quad \quad -\sum_\mathbf{q} \frac{\lambda(\mathbf{q})}{2}\left(\hat{u}^x_{-\mathbf{q}}\hat{p}^y_\mathbf{q}- \hat{u}^y_{-\mathbf{q}}\hat{p}^x_\mathbf{q}\right),
\end{eqnarray}
where 
 $\hat{\mathbf u}_\mathbf{q}\equiv (\hat{u}^x_\mathbf{q}, \hat{u}^y_\mathbf{q})^T,
 \hat{\mathbf p}_\mathbf{q}\equiv (\hat{p}^x_\mathbf{q}, \hat{p}^y_\mathbf{q})^T$ and 
 \begin{equation}
 [\hat{u}^\alpha_\mathbf{q}, \hat{p}^\beta_{\mathbf{q}'}]=i\delta_{\alpha\beta}\delta_{\mathbf{q},-\mathbf{q}'},
\  [\hat{u}^\alpha_\mathbf{q}, \hat{u}^\beta_{\mathbf{q}'}]=[\hat{p}^\alpha_\mathbf{q}, \hat{p}^\beta_{\mathbf{q}'}]=0.
\end{equation}
With the transformation
\begin{equation}
 \hat{u}^\alpha_\mathbf{q}=\frac{1}{\sqrt{2\omega_\mathbf{q}}}\left(\hat{a}_{\mathbf{q},\alpha}+\hat{a}^\dagger_{-\mathbf{q},\alpha}\right),\
\hat{p}^\alpha_\mathbf{q}=-i\sqrt{\frac{\omega_\mathbf{q}}{2}}\left(\hat{a}_{\mathbf{q},\alpha}-\hat{a}^\dagger_{-\mathbf{q},\alpha}\right),
\end{equation}
 Hamiltonian Eq.(\ref{H_k_phonon}) becomes
\begin{eqnarray}
\hat{H}_{\rm ph} &=&\sum_\mathbf{q} \omega_\mathbf{q}\left(\hat{a}^\dagger_{\mathbf{q},x}\hat{a}_{\mathbf{q},x}+\hat{a}^\dagger_{\mathbf{q},y}\hat{a}_{\mathbf{q},y}+1\right) \nonumber\\
&& \quad \ \ +\sum_\mathbf{q} \frac{i\lambda_{\mathrm{ph}}(\mathbf{q})}{2}\left(\hat{a}^\dagger_{\mathbf{q},x}\hat{a}_{\mathbf{q},y}-\hat{a}^\dagger_{\mathbf{q},y}\hat{a}_{\mathbf{q},x}\right),
\end{eqnarray}
which can be diagonalized as 
\begin{equation}
\hat{H}_{\rm ph}=\sum_{\mathbf{q},\chi=\pm}\omega_{\mathbf{q},\chi}\left(\hat{b}^\dagger_{\mathbf{q}\chi}\hat{b}_{\mathbf{q}\chi}+\frac{1}{2}\right),
\end{equation}
where 
\begin{equation}
\hat{b}_{\mathbf{q}\chi}=\frac{1}{\sqrt{2}}(\hat{a}_{\mathbf{q},x}+i\chi \hat{a}_{\mathbf{q},y}),\ \chi=\pm 1,
\end{equation}
are two chiral phonon operators with frequency $\omega_{\mathbf{q},\chi}=\omega_\mathbf{q}+ \chi \lambda(\mathbf{q})/2$ and chirality $\chi=\pm 1$. We assume $\omega_\mathbf{q}=c q$ for acoustic mode and the chiral phonon splitting $\lambda(\mathbf{q})=\lambda q$, where $c, \lambda$ are constants. This is consistent with the result in \cite{ZhangNiu2015} that the chiral phonon spliting  is zero at  $\Gamma$ point of the phonon BZ due to the TRS at this point and is finite at  $K, K'$ points due to TRS breaking at these points.

Denoting the polarization vector of the chiral phonons as $\boldsymbol{\epsilon}_\chi=\frac{1}{\sqrt{2}}\begin{pmatrix}1\\ - i \chi\end{pmatrix}$, the displacement field can be written as 
\begin{equation}
 \hat{\mathbf u}_\mathbf{q}=\frac{1}{\sqrt{2\omega_\mathbf{q}}}
 \sum_{\chi=\pm1}
 \left(\boldsymbol{\epsilon}_\chi \hat{b}_{\mathbf{q}\chi}+\boldsymbol{\epsilon}_\chi^* \hat{b}^\dagger_{-\mathbf{q}\chi}\right).
\label{eq:phonon-displacement}
\end{equation}
Its Green's function (GF) $\hat{D}(\tau, \mathbf{q})\equiv\langle T_\tau \hat{\mathbf{u}}_q(\tau) \hat{\mathbf{u}}^\dag_{q}(0)\rangle$ in the chiral phonon basis with Matsubara frequency is equal to 
\begin{equation}\label{eq:chiral-Matsubara-D}
 \hat{D}(i\omega_n, \mathbf{q})
=\frac{1}{2\omega_\mathbf{q}}\sum_{r, \chi=\pm 1}
\frac{r |\chi\rangle\langle \chi|}{i\omega_n-r\omega_{\mathbf{q} \chi}}
\end{equation}
where $|\chi\rangle, \chi=\pm 1$ are the chiral phonon eigenstates.

In this work we consider only the in-plane acoustic phonon modes of the monolayer TMDs and ignore the intervalley scatterings due to e-phonon interaction. 
The electron-acoustic-phonon coupling in monolayer TMDs can be obtained from the elastic theory and includes a deformation potential (DP) and a pseudogauge vector potential (VP)~\cite{Guinea2014, Shan2020, Shan2022}, i.e., 
\begin{equation}\label{e_ph}
\hat{H}_{\rm e\text{-}ph}=\hat{H}^d_{\rm e\text{-}ph}+ \hat{H}^p_{\rm e\text{-}ph}, 
\end{equation}
where the DP and VP part are respectively
\begin{equation}\label{e_ph_D_P}
\hat{H}^{d/p}_{\rm e\text{-}ph}=\sum_{\mathbf{k, q}, \tau}\psi^\dag_{\mathbf{k+q}, \tau}[\mathbf{\hat{u}}_{\mathbf{q}}\cdot \mathbf{T}^{d/p}_\tau(\mathbf{q})]\psi_{\mathbf{k}, \tau},
\end{equation}
with~\cite{Guinea2014, Shan2020, Shan2022}
\begin{eqnarray}
\mathbf{T}^{d}_\tau(\mathbf{q})&=&i g_d \mathbf{q}, \nonumber\\
\mathbf{T}^{p}_{\tau=-1}(\mathbf{q}) &=& ig_p[\mathbf{q}\cdot\boldsymbol{\sigma}, (\mathbf{q}\cross\boldsymbol{\sigma})_z], \nonumber\\
\mathbf{T}^{p}_{\tau=1}(\mathbf{q})&=&{\cal K}[\mathbf{T}^{p}_{\tau=-1}(\mathbf{-q})].
\end{eqnarray}
Here ${\cal K}$ is the complex conjugate operator. The DP is independent of the valley index whereas the VP force $\mathbf{T}^{p}_{\tau}(\mathbf{q})$ in the two valleys are related by the TRS.

\section{Chiral phonon induced electron anomalous Hall effect}\label{AHE}

We next compute the electron anomalous Hall effect due to the chiral phonon interaction with the electrons in the monolayer TMD. We assume that the electron Fermi surface locates at the topmost valence band as shown in Fig.\ref{fig:band}. The spin is then polarized on the Fermi surface in each valley with opposite directions in the two valleys. 

\begin{figure}[t]  
   \includegraphics[width=0.47\textwidth]{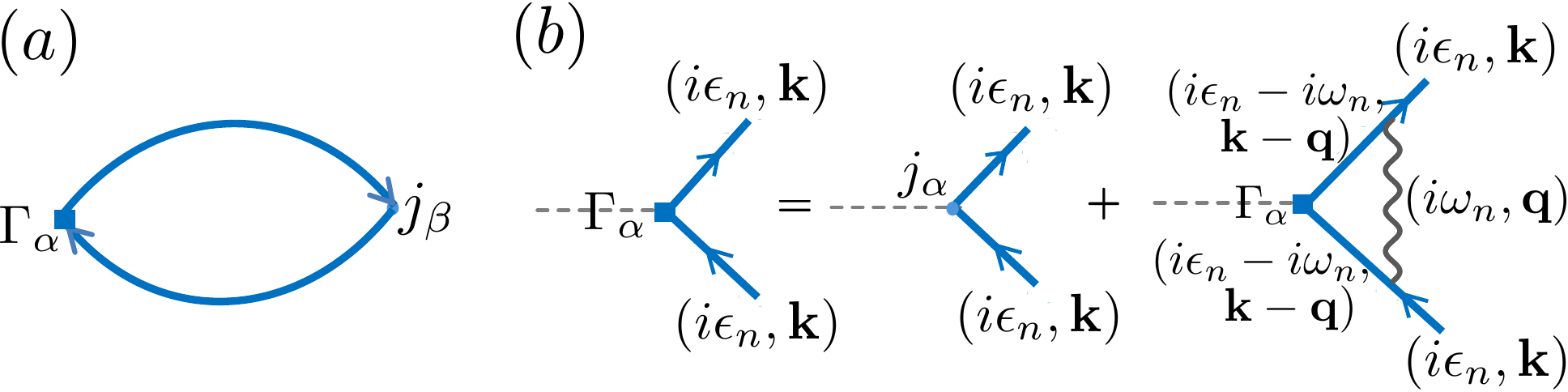} 
 \caption{  \justifying (a) The Feynman diagram of the Hall conductivity from the linear response theory. (b)The Feynman diagram describing the current vertex renormalization equation. }
\label{AH_conductivity} 
\end{figure}

The electron-phonon Hamiltonian in Eq.(\ref{e_ph}) can be written in the chiral phonon basis  as
\begin{eqnarray}
H_{e\text{-}ph}
&&=\sum_{k,\mathbf{q},\tau,s, \chi}\frac{1}{\sqrt{2\omega_{\mathbf{q}}}} \nonumber\\
&& \psi^\dagger_{\mathbf{k+q},\tau, s}
\left[\hat{M}_{+\chi,\tau}(\mathbf{q})b_{\mathbf{q}, \chi}
+\hat{M}_{-\chi,\tau}(\mathbf{q})b^\dagger_{-\mathbf{q},\chi}\right]
 \psi_{\mathbf{k},\tau, s}, \nonumber\\
\label{eq:eph-chiral}
\end{eqnarray}
where the first and second term correspond to absorbing and emitting a chiral phonon during the e-phonon scattering process respectively. The interaction vertex 
\begin{equation}
\hat{M}_{r,\chi,\tau}(\mathbf{q})
=\frac{\hat{T}^x_\tau(\mathbf{q})-ir\chi \hat{T}^y_\tau(\mathbf{q})}{\sqrt2},\ r, \chi, \tau=\pm 1,
\label{eq:Mrchi-definition}  
\end{equation}
with 
\begin{eqnarray}
&\hat{T}^x_\tau(\mathbf{q})=ig_dq_x \hat{I}_2+ig_p(q_x\hat{\sigma}_x-\tau q_y\hat{\sigma}_y), \\
&\hat{T}^y_\tau(\mathbf{q})=ig_dq_y\hat{I}_2+ig_p(-q_y\hat{\sigma}_x-\tau q_x\hat{\sigma}_y).
\end{eqnarray}

The Hall conductivity of electrons can be obtained from the Kubo-Streda formula~\cite{Streda1982}, which includes a Fermi surface contribution and a Fermi sea contribution. The latter is insensitive to the the electron-phonon scatterings~\cite{Sinitsyn2005, Sinitsyn2007} and vanishes for the electron Hamiltonian with TRS. We then focus on the Fermi surface contribution in this work. The Hall conductivity from the Fermi surface in the dc limit can be obtained from the Feynman diagram in 
Fig.\ref{AH_conductivity}(a) as~\cite{Streda1982, Sinitsyn2007}
\begin{eqnarray}\label{eq:Hall_conductivity}
\sigma_{xy}
&=&\int_{-\infty}^{\infty}\frac{d\epsilon}{2\pi}
\int\frac{d^2 \mathbf{k}}{(2\pi)^2} [-\partial_\epsilon n_F(\epsilon)] \nonumber\\
&&\operatorname{Tr}
[\hat \Gamma_x(\epsilon, \mathbf{k}) \hat{G}^R(\epsilon, \mathbf{k})\hat j_y \hat{G}^A(\epsilon, \mathbf{k})],
\label{eq:Kubo-formula}
\end{eqnarray}
where $\hat{G}^{R/A}(\epsilon, \mathbf{k})$ is the renormalized retarded (advanced) GF of the electrons and 
the bare current vertex 
\begin{equation}
\hat j_x=e\,v\tau \hat{\sigma}_x,
\qquad
\hat j_y=e\,v \hat{\sigma}_y.
\end{equation}

The renormalized current vertex $\hat{\Gamma}_x$ in the imaginary time space  satisfies the iterative equation described by the Feynman diagram in Fig.\ref{AH_conductivity}(b) as~\cite{Chen2023, ZhangChen2024} 
\begin{eqnarray}\label{eq:vertex_recursion}
&&\hat{\Gamma}_x(i\epsilon_n,\mathbf{k})=\hat{j}_x +\sum_{r, \chi,\tau}\sum_{i\omega_n, \mathbf{q}} \nonumber\\
&&\quad
\hat{M}_{r,\chi,\tau}(\mathbf{q})\hat{G}(i\epsilon_n-i\omega_n, \mathbf{k}-\mathbf{q})\hat{\Gamma}_x(i\epsilon_n-i\omega_n, \mathbf{k}-\mathbf{q}) \nonumber\\
&&\quad \quad \hat{G}(i\epsilon_n-i\omega_n, \mathbf{k}-\mathbf{q})\hat{M}^\dag_{r, \chi, \tau}(\mathbf{q})D_{r \chi}(i\omega_n, \mathbf{q}),
\end{eqnarray}
where 
 \begin{equation}
D_{r\chi}(i\omega_n, \mathbf{q})
=\frac{1}{2\omega_\mathbf{q}}
\frac{r}{i\omega_n-r\omega_{\mathbf{q}\chi}}
\label{eq:chiral-pole-D}
\end{equation}
 is  the chiral phonon propagator and $\hat{G}(i\epsilon_n, \mathbf{k})$ is the electron GF in the Matsubara frequency space. The renormalized current vertex $\hat{\Gamma}_x$ in Eq.(\ref{eq:Hall_conductivity}) corresponds to the one in Eq.(\ref{eq:vertex_recursion}) analytically continued to the real energy space.

\subsection{Electron Green's function}

We assume the Fermi energy $|\epsilon_F|\gg T$ for hole doping in the system
so the factor $-\partial_\epsilon n_F(\epsilon)\sim\delta(\epsilon-\epsilon_F)$ and the Hall conductivity in Eq.(\ref{eq:Kubo-formula})  comes from the electrons on the Fermi surface.
In the following calculation, we assume by default that $(\epsilon, \mathbf{k})$ take values on the Fermi surface that locates on the topmost valence band shown in Fig.\ref{fig:band}.

The bare electron GF with Matsubara frequency is 
\begin{equation}
G^0_{\tau s}(i\epsilon_n, \mathbf k)
=\sum_{\eta=\pm1}
\frac{P_{\tau s,\eta}(\mathbf k)}
{i\epsilon_n-\varepsilon_{\tau s}-\eta E_{\tau s}(\mathbf k)},
\end{equation}
where the band projection operator 
\begin{eqnarray}
&& P_{\tau s,\eta}(\mathbf k)=|\eta\rangle\langle \eta| \nonumber\\
&&=\frac12\left[
1+\frac{\eta}{E_{\tau s}(\mathbf{k})}
\left(v\tau k_x\sigma_x+v k_y\sigma_y+m_{\tau s}\sigma_z\right)
\right].
\label{eq:band-projector}
\end{eqnarray}

\begin{figure}[t]  
   \includegraphics[width=0.45\textwidth]{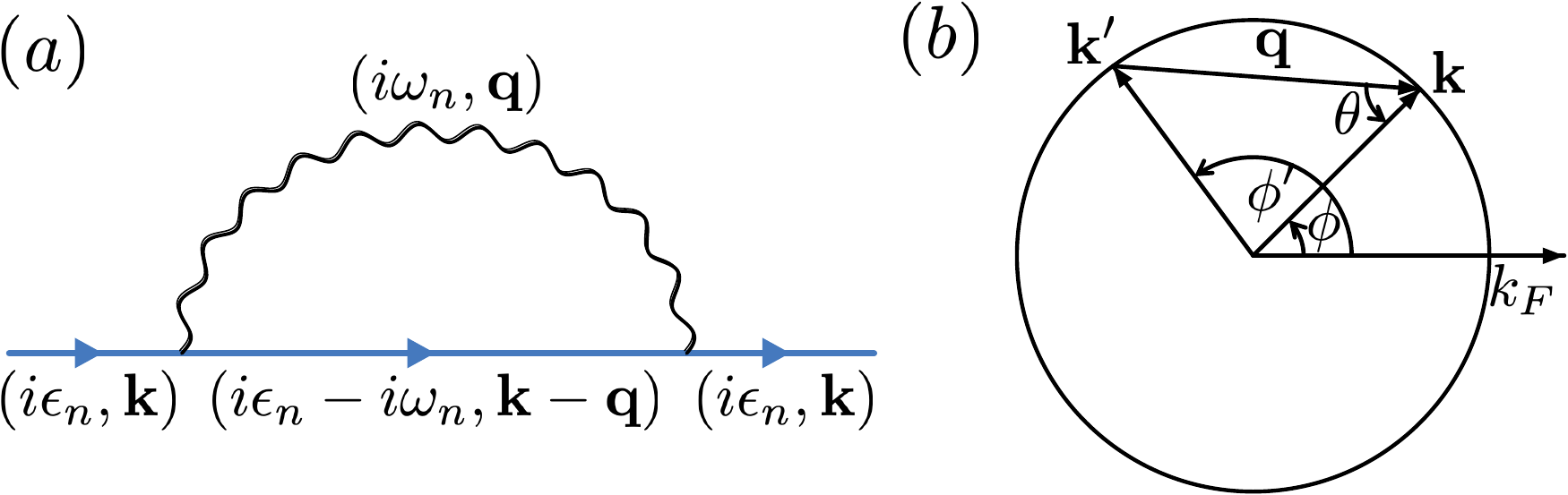} 
 \caption{  \justifying (a)The Feynman diagram of the electron self-energy due to the e-phonon interaction. The straight line represents the bare electron GF and the curved line represent the phonon propagator. (b)The e-phonon scattering process near the Fermi surface. The initial electron momentum $\mathbf{k}$ is on the Fermi surface. The final electron momentum $\mathbf{k}'=\mathbf{k}-\mathbf{q}$ where $\mathbf{q}$ is the phonon momentum. Since the phonon energy $\omega_{\mathbf{q}, \chi}\ll \epsilon_F$, $\mathbf{k}'$ is near the Fermi surface and $\cos\theta\approx q/2k_F$.}
 \label{self_energy} 
\end{figure}

The leading order AH conductivity corresponds to the renormalized electron GF under the first Born approximation~\cite{Sinitsyn2007}
\begin{equation}
\hat{G}(i\epsilon_n, \mathbf{k})=[\hat{G}^{-1}_0(i\epsilon_n, \mathbf{k})-\Sigma(i\epsilon_n, \mathbf{k})]^{-1},
\end{equation}
where the electron self-energy  due to the electron-phonon interaction  for a given valley and spin polarization is described by the Feynman diagram in Fig.\ref{self_energy}(a) as 
\begin{eqnarray}\label{eq:Fock-start}
&&\Sigma_{\tau s}(i\epsilon_n, \mathbf k)
=-\frac{1}{\beta L^2}
\sum_{i\omega_n, \mathbf q}\sum_{r=\pm1}\sum_{\chi=\pm1} \nonumber\\
&&M_{r,\chi,\tau}(\mathbf q)
G^0_{\tau s}(i\epsilon_n-i\omega_n, \mathbf k-\mathbf q)
M^\dagger_{r,\chi,\tau}(\mathbf q)
D_{r\chi}(i\omega_n, \mathbf q). \nonumber\\
\end{eqnarray}

The real part of the self-energy only results in a shift of the electron energy so we neglect it. 
The imaginary part of the electron self-energy analytically continued to real energy space after the sum of the phonon Matsubara frequency $i\omega_n$ becomes
\begin{align}
&-\operatorname{Im}\Sigma^R_{\tau s}(i\epsilon_n\to \epsilon+i0^+, \mathbf k)
=\pi\sum_{\chi,\eta}
\int\frac{d^2 \mathbf{q}}{(2\pi)^2}\frac{1}{2\omega_\mathbf{q}}\notag\\
&
 \Bigg\{
\Big[n_B(\omega_{\mathbf{q}\chi})+1
-n_F( \epsilon_{\tau s \eta}(\mathbf{k}))\Big] 
\delta\!\left[\epsilon- \epsilon_{\tau s \eta}(\mathbf{k})-\omega_{\mathbf{q}\chi}\right]
\mathcal S_{+,\chi,\tau,\eta}
\notag\\
&\  +
\Big[n_B(\omega_{\mathbf{q}\chi})
+n_F( \epsilon_{\tau s \eta}(\mathbf{k}))\Big]
\delta\!\left[\epsilon- \epsilon_{\tau s \eta}(\mathbf{k})+\omega_{\mathbf{q}\chi}\right]
\mathcal S_{-,\chi,\tau,\eta}
\Bigg\},
\label{eq:SE}
\end{align}
where the first and second terms correspond to absorbing and emitting a phonon respectively and 
\begin{equation}
\mathcal S_{r,\chi,\tau,\eta}
\equiv
M_{r,\chi,\tau}(\mathbf{q})P_{\tau s,\eta}(\mathbf{k-q}) M^{\dagger}_{r,\chi,\tau}(\mathbf{q}).
\label{eq:channel-diagonal-kernel}
\end{equation}

The DP and pseudogauge VP part of $\mathcal S_{r,\chi,\tau, \eta}$ are respectively
\begin{eqnarray}
&\mathcal S^{\rm DP}_{r,\chi,\tau, \eta}
=\frac{g_d^2 q^2}{4}\left[
1+\frac{\eta}{E_{\tau s}(\mathbf{k}')}
\left(v\tau k'_x\sigma_x+v k'_y\sigma_y+m_{\tau s}\sigma_z\right)
\right], \nonumber\\
&\mathcal S^{\rm VP}_{r,\chi,\tau, \eta}
=\frac{g_p^2 q^2}{2} \left[1-\frac{r\chi\tau\eta m_{\tau s}}{E_{\tau s}(\mathbf{k}')}\right]
(1+r\chi\tau\sigma_z), \notag
\end{eqnarray}
where $\mathbf{k}'=\mathbf{k}-\mathbf{q}$.

We further assume that the chiral phonon frequency $\omega_{q,\chi}\ll |\epsilon_F|$, so the electron
after absorbing or emitting a phonon still locates near the Fermi surface, as shown in Fig.\ref{self_energy}(b). 
 Denote 
\begin{equation}
\mathbf k=k(\cos\phi,\sin\phi),\ \ 
\mathbf q=q\big(\cos(\phi+\theta),\sin(\phi+\theta)\big),
\end{equation}
the momentum after scattering can be expressed as 
\begin{equation}\label{eq:k'}
\mathbf{k}'\approx k(\cos\phi', \sin\phi'),\ \phi'=\phi+\pi+2 \theta
\end{equation}
with $\cos{\theta}\approx q/2k$. Integrating out the angular part of $\mathbf{q}$ in the self-energy Eq.(\ref{eq:SE}), we get the imaginary part of the 
electron self-energy
 due to the DP as 
\begin{equation}\label{DP_self_energy}
-\operatorname{Im}\Sigma^{\rm R, DP}_{\tau s}(\epsilon, \mathbf{k}))
=g^2_d[\xi^d_{0}
+\xi^d_{1}(\tau\cos\phi\,\sigma_x+\sin\phi\,\sigma_y)
+\xi^d_{z}\sigma_z],
\end{equation}
with the coefficients
\begin{align}
\xi^d_{0}
&=\sum_{\chi=\pm 1}\int_0^{2k}\frac{q^2\,dq}{8\pi v^2 \omega_\mathbf{q}}\frac{E_{\tau s}(\mathbf{k})}{ \sqrt{k^2-q^2/4}}F(\frac{\omega_{\mathbf{q}\chi}}{T}, \frac{\epsilon}{T}),\\
\xi^d_{1}
&=-\sum_{\chi=\pm 1}\int_0^{2k}\frac{q^2\,dq}{8\pi v \omega_\mathbf{q}}
(1-\frac{q^2}{2k^2})\frac{1}{\sqrt{1-\frac{q^2}{4k^2}}}F(\frac{\omega_{\mathbf{q}\chi}}{T}, \frac{\epsilon}{T}),\\
\xi^d_{z}
&=-\sum_{\chi=\pm 1}\int_0^{2k}\frac{q^2\,dq}{8\pi v^2\omega_\mathbf{q}}
\frac{m_{\tau s}}{\sqrt{k^2-q^2/4}}F(\frac{\omega_{\mathbf{q}\chi}}{T}, \frac{\epsilon}{T}), \label{DP_SE_coefficients}
\end{align}
where $F(\frac{\omega_{\mathbf{q}\chi}}{T}, \frac{\epsilon}{T})\equiv n_B(\omega_{\mathbf{q}\chi})+n_F(\epsilon+\omega_{\mathbf{q}\chi})$ with $n_B$ and $n_F$ the  distribution functions of Bosons and Fermions respectively. We can see that the electron self-energy due to DP scatterings is independent of the valley index because the DP is the same in the two valleys.

For convenience, we introduce a variable $x=q/2k$ and write $\xi^d_{0}, \xi^d_{1}, \xi^d_{z}$ as 
\begin{equation}
\xi^d_{0}=\frac{\gamma_0}{2} \tilde{\xi}^d_{0},\  
\xi^d_{1}=-\frac{\kappa}{2}  \gamma_0\tilde{\xi}^d_{1},\    
\xi^d_{z}=-\frac{\tilde{m}}{2}\gamma_0 \tilde{\xi}^d_{z}, \ 
\end{equation}
where
\begin{align}
 \gamma_0\equiv \frac{2E_{\tau s}(\mathbf{k}) k^2}{\pi v^2 \Omega_{BG}}, \ \Omega_{BG}\equiv 2c k,
 \ \kappa\equiv \frac{vk }{E_{\tau s}(\mathbf{k})},
 \tilde{m}\equiv \frac{m_{\tau s} }{E_{\tau, s}(\mathbf{k})} \notag
\end{align}
 and 
$\tilde{\xi}^d_{0}, \tilde{\xi}^d_{1}, \tilde{\xi}^d_{z}$ are dimensionless constants as
\begin{align}
\tilde{\xi}^d_{0}=\sum_{\chi=\pm 1}\int_0^{1} 
   \frac{x^2}{w(x)\sqrt{1-x^2}}F(\frac{\omega_{\mathbf{q},\chi}}{T}, \frac{\epsilon}{T}), \\
\tilde{\xi}^d_{1}=\sum_{\chi=\pm 1}\int_0^{1} 
   \frac{x^2(1-2x^2)}{w(x)\sqrt{1-x^2}}F(\frac{\omega_{\mathbf{q},\chi}}{T}, \frac{\epsilon}{T}), \\
\tilde{\xi}^d_{z}=\sum_{\chi=\pm 1}\int_0^{1} 
   \frac{x^2}{w(x)\sqrt{1-x^2}}F(\frac{\omega_{\mathbf{q},\chi}}{T}, \frac{\epsilon}{T}),   
\end{align}
with $w(x)\equiv \omega_\mathbf{q}/\Omega_{BG}$.

The imaginary part of the electron self-energy due to the pseudogauge VP is
\begin{equation}\label{VP_self_energy}
-\operatorname{Im}\Sigma^{\rm R, VP}_{\tau s}(\epsilon, \mathbf{k}))=g^2_p[\xi^p_{0} +\xi^p_{z}\sigma_z],
\end{equation}
with the coefficients
\begin{align}
\xi^p_{0}
&=\sum_\chi\int_0^{2k}\frac{q^2\,dq}{4\pi v^2 \omega_\mathbf{q}}\frac{E_{\tau, s}(\mathbf{k})}{ \sqrt{k^2-q^2/4}}F(\frac{\omega_{\mathbf{q},\chi}}{T}, \frac{\epsilon}{T}), \label{VP_SE_coef_0}\\
\xi^p_{z}
&=\sum_\chi\int_0^{2k}\frac{q^2\,dq}{4\pi v^2\omega_\mathbf{q}}
\frac{\chi\tau \omega_{\mathbf{q},\chi}+m_{\tau s}}{\sqrt{k^2-q^2/4}}
F(\frac{\omega_{\mathbf{q},\chi}}{T}, \frac{\epsilon}{T}), \label{VP_SE_coef_z}
\end{align}
which can be written as 
\begin{equation}
\xi^p_{0}=\gamma_0 \tilde{\xi}^p_{0}, \quad \xi^p_{z}=\gamma_0 \tilde{\xi}^p_{z}
\end{equation}
with the dimensionless constants
\begin{align}
\tilde{\xi}^p_{0}&=  \tilde{\xi}^d_{0},  \\
 \tilde{\xi}^p_{z}&= \sum_{\chi=\pm 1}\int_0^{1} 
   \frac{x^2}{w(x)\sqrt{1-x^2}}F(\frac{\omega_{\mathbf{q},\chi}}{T}, \frac{\epsilon}{T})
   [\tilde{m}+\frac{\tau \chi \omega_{\mathbf{q}, \chi}}{E_{\tau, s}(\mathbf{k})}]. \label{SE_VP}
\end{align}
In contrast to the DP, we can see that the electron self-energy due to VP scatterings is different in the two valleys because the VP is different in the two valleys. 

The retarded and advanced electron GF under the first Born approximation is then 
\begin{align}
& \hat{G}^{R/A}_{\tau s}(\epsilon, \mathbf{k})=\hat{G}(i\epsilon_n\to \epsilon\pm i 0^+, \mathbf{k})\notag\\
=&\frac{1}{\mathcal D_{R/A}(\epsilon, \mathbf{k})}\Big[(\epsilon-\epsilon_{\tau s}\pm i\Sigma_0)I_2 +(m_{\tau s}\mp i\Sigma_z)\sigma_z
\notag\\
&\quad \quad \ \ +(vk\mp i\Sigma_1)(\tau\cos\phi\,\sigma_x+\sin\phi\,\sigma_y)\Big],
\label{eq:GR-explicit-v5}
\end{align}
where 
\begin{equation}
\Sigma_0=g_d^2\xi^d_{0}+g_p^2\xi^p_{0},\
\Sigma_1=g_d^2\xi^d_{1},\
\Sigma_z=g_d^2\xi^d_{z}+g_p^2\xi^p_{z}
\label{eq:Sigma-components-v5}
\end{equation}
and 
\begin{equation}
\mathcal D_{R/A}(\epsilon, \mathbf{k})=[\epsilon-\epsilon_{\tau s +}(\mathbf{k})\pm i \gamma_+][\epsilon-\epsilon_{\tau s -}(\mathbf{k})\pm i \gamma_-],
\label{eq:DRDA-v5}
\end{equation}
with
\begin{align}
\gamma_{\pm}
=\Sigma_0\pm \kappa\Sigma_1\pm \tilde{m}\Sigma_z.
\label{eq:gamma-v5}
\end{align}
For brevity, we have dropped the variable $(\epsilon, \mathbf{k})$ of $\Sigma_{0, 1, z}$ and $\gamma_\pm$ in the above equations.

\subsection{Current vertex renormalization}

We next study the current vertex renormalization due to the e-phonon interaction.
The renormalization Eq.(\ref{eq:vertex_recursion}) gives two main effects~\cite{Sinitsyn2007}.
 One is an enhancement of the longitudinal component of the bare current vertex $\hat{j}_x$. The other is the generation of a transverse current  which in turn contributes to the anomalous Hall current. The first effect involves only electron scatterings within the Fermi surface band, whereas the second effect involves interband electron scatterings~\cite{Sinitsyn2007}.

 For the isotropic case we study in this work, the renormalization of the longitudinal current component can be solved from Eq.(\ref{eq:vertex_recursion}) projected to the Fermi surface band~\cite{Sinitsyn2007, ZhangChen2024}, which we assumed to be the valence band with $\eta=-1$ in this work. The projected recursion equation of the band diagonal current in valley $\tau$ is 
\begin{eqnarray}\label{renormalized_current}
&&\Gamma^{--}_x(\epsilon, \mathbf{k})= j_x^{--}(\mathbf{k}) + \sum_{r, \chi}\int \frac{d^2\mathbf{q}}{(2\pi)^2}\frac{1}{2\omega_\mathbf{q}} \nonumber\\
&&\quad
\frac{|_\mathbf{k}\langle -|M_{r, \chi, \tau}|-\rangle_{\mathbf{k}'}|^2}{[E_{\tau s}(\mathbf{k}')-E_{\tau s}(\mathbf{k})-r\omega_{\mathbf{q}, \chi}]^2+\gamma_{-}^2(\epsilon, \mathbf{k})}\Gamma^{--}_x(\epsilon',\mathbf{k}'), \nonumber\\
\end{eqnarray}
where $|-\rangle_\mathbf{k} = \begin{pmatrix}
-\sqrt{(1-\frac{m_{\tau s}}{E_{\tau s}(\mathbf{k})})/2}\\
\tau e^{i\tau\phi}\sqrt{(1+\frac{m_{\tau s}}{E_{\tau s}(\mathbf{k})})/2}
\end{pmatrix}$ 
is the electron eigenstate of the  band $\eta=-1$ with energy $\epsilon_{\tau s -}(\mathbf{k})$ and momentum $\mathbf{k}=k(\cos\phi, \sin\phi)$. 
The band diagonal current  $\Gamma^{--}_x(\epsilon, \mathbf{k})\equiv \ _\mathbf{k}\langle -|\hat{\Gamma}_x|-\rangle_\mathbf{k},\ j_x^{--}(\mathbf{k})\equiv\ _\mathbf{k}\langle - |\hat{j}_x|-\rangle_\mathbf{k}=ev\kappa \cos\phi$. 
 For small $\gamma_-(\epsilon, \mathbf{k})$, 
\begin{eqnarray}
&&\frac{1}{[E_{\tau s}(\mathbf{k}')-E_{\tau s}(\mathbf{k})-r\omega_{\mathbf{q}, \chi}]^2+\gamma_-^2(\mathbf{k})}\nonumber\\
&& \approx
\frac{\pi}{\gamma_-(\epsilon, \mathbf{k})}\delta[E_{\tau s}(\mathbf{k}')-E_{\tau s}(\mathbf{k})-r\omega_{\mathbf{q}, \chi}].
\label{eq:pinching-v5}
\end{eqnarray}
The electron momentum after e-phonon scattering is $\mathbf{k}'=\mathbf{k}-\mathbf{q}\approx k(\cos \phi', \sin\phi')$ and the energy $E_{\tau s}(\mathbf{k}')=E_{\tau s}(\mathbf{k})+r\omega_{\mathbf{q}, \chi}$.

\begin{figure}[t]  
   \includegraphics[width=0.45\textwidth]{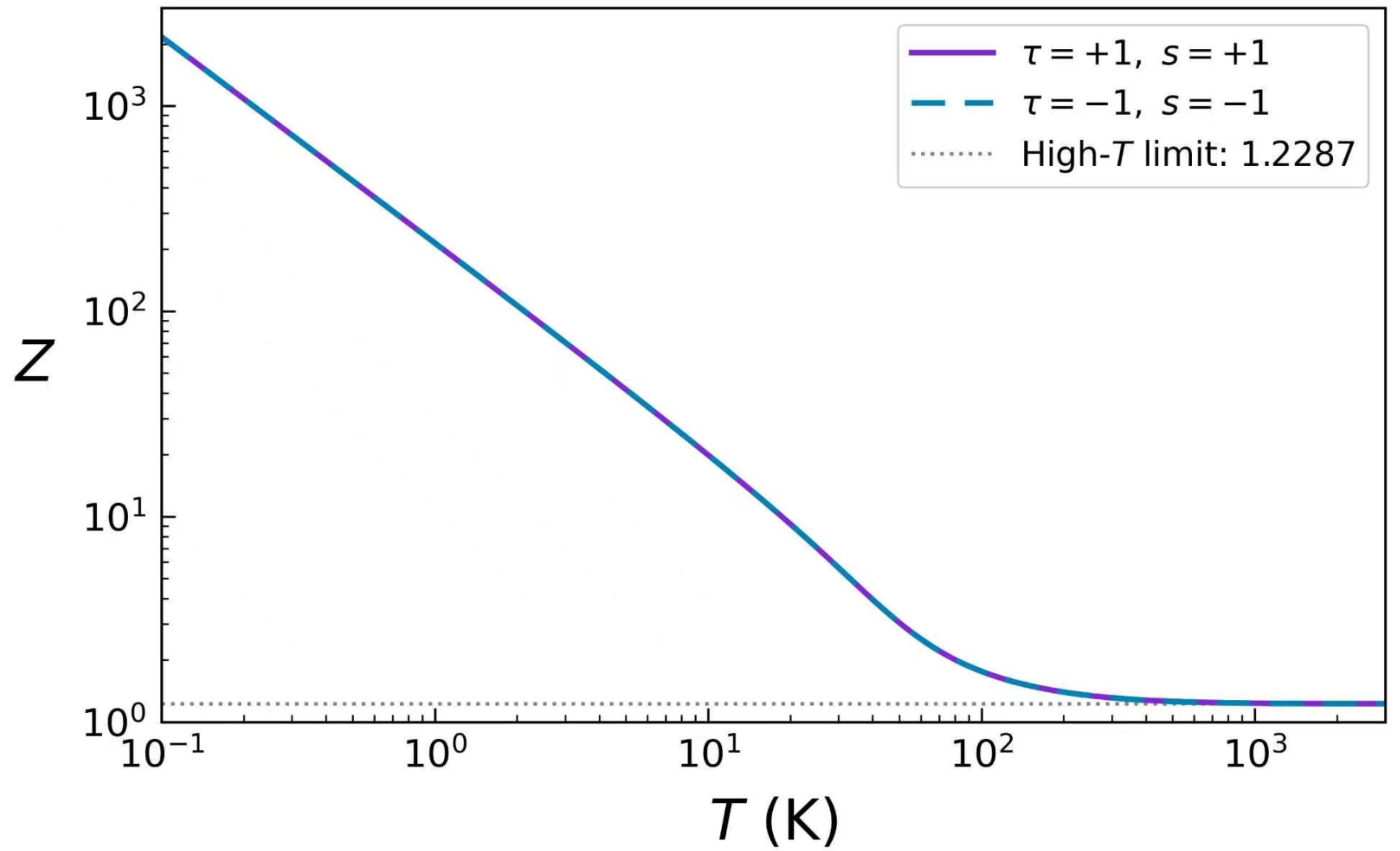} 
 \caption{  \justifying The current vertex  renormalization factor $Z$ at each valley as a function of temperature $T$ at $\epsilon_F=-0.8eV, \Delta=1.66{\rm eV}, \lambda_{SO}=0.075{\rm eV}, v=3.51 {\rm eV \mathring{A}}, c=10^4 m/s, \lambda=0.06 {\rm eV \mathring{A}}, \rho=0.2$. The data comes from Ref.\cite{Xiao2012} for monolayer $\rm Mo S_2$. $Z$ has a very weak dependence on the valley index $\tau$ from the VP scatterings at finite temperature. At $T \to \infty$, $Z$ approaches a constant.}
 \label{fig:Z} 
\end{figure} 

Taking the approximation 
of $\mathbf{k}'$ in Eq.(\ref{eq:k'}), the matrix element in Eq.(\ref{renormalized_current}) up to the first order of $\omega_{\mathbf{q},\chi}/E_{\tau s}(\mathbf{k})$ is 
\begin{eqnarray}
\sum_{r=\pm1}| _\mathbf{k}\langle -|M^{(d)}_{r\chi,\tau}|-\rangle_\mathbf{k'}|^2
&\approx & g_d^2 q^2 \left[1-\frac{v^2 q^2}{4E^2_{\tau s}(\mathbf{k})}\right], \nonumber\\
\sum_{r=\pm1}| _\mathbf{k}\langle -|M^{(p)}_{r\chi,\tau}|-\rangle_\mathbf{k'}|^2
&\approx & g_p^2 q^2 \frac{v^2 k^2}{E_{\tau s}^2(\mathbf{k})} \nonumber
\end{eqnarray}
for DP and VP respectively.
The renormalized longitudinal band-diagonal current $\Gamma^{--}_x$ can be solved by iteration of Eq.(\ref{renormalized_current}) and we get
\begin{equation}
\Gamma^{--}_x(\epsilon, \mathbf{k})=Z(\epsilon, k) j_x^{-}(\mathbf{k}), 
\end{equation}
with the renormalization factor 
\begin{equation}\label{Z}
Z(\epsilon, k)=\frac{b_0}{b_0-b_1}
\end{equation}
where $b_0, b_1$ can be written as dimensionless constants up to the first order of $\omega_{\mathbf{q}, \chi}$ as 
\begin{eqnarray}
 && b_0 (\epsilon, k)=\gamma_-(\epsilon, \mathbf{k})/\gamma_0, \label{b_0}\\
 && b_1 (\epsilon, k)=\sum_{\chi=\pm 1}\int_0^{1} 
   \frac{x^2}{w(x)\sqrt{1-x^2}}F\left(\frac{\omega_{\mathbf{q},\chi}}{T}, \frac{\epsilon}{T}\right) \nonumber\\
  &&\ \ \left\{1-\frac{v^2 k^2}{E_{\tau s}^2(\mathbf{k})} x^2+ \rho^2 \left[\frac{v^2 k^2}{E_{\tau s}^2(\mathbf{k})} - \tau \chi \frac{m_{\tau s} \omega_{\mathbf{q},\chi}}{E_{\tau s}^2(\mathbf{k})} \right] \right\} \label{b_1} \nonumber\\
\end{eqnarray}
where $\rho \equiv g_d/g_p$.


The renormalization factor $Z$ is dependent on the ratio $\rho$ of the DP and VP coupling constants, but not their absolute values. The temperature dependence of $Z$ for given Fermi energy, phonon frequency  and $\rho$ is plotted in Fig.\ref{fig:Z}. There is a very weak valley dependence of $Z$ due to the VP scattering as shown in Fig.\ref{fig:Z}. If one ignores the frequency difference of the left and right chiral phonons, the renormalization factor in Eq.(\ref{fig:Z}) with only DP contribution reduces to that of the linear phonon scatterings in \cite{ZhangChen2024}.
For both linear and chiral phonon scatterings, at high temperature $T\gg T_{BG}\equiv 2 c k_F$ the renormalization factor $Z$  due to  DP scatterings reduces to that of the Gaussian disorder shown in \cite{Sinitsyn2007}.


\subsection{The electron anomalous  Hall conductivity}

\begin{figure}[t]  
   \includegraphics[width=0.48\textwidth]{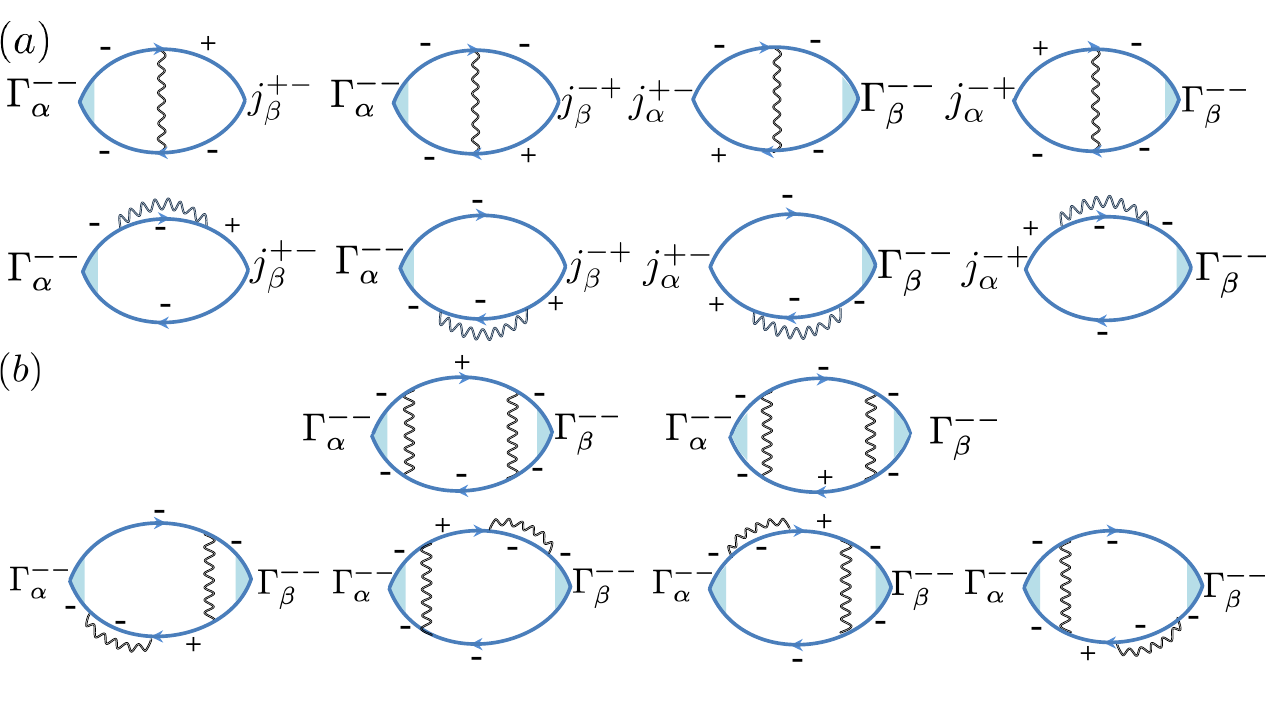} 
 \caption{\justifying The Feynman diagrams of the extrinsic contribution to the AHE in the eigenband basis. (a)The Feynman diagrams for the side jump contribution. (b)The Feynman diagrams for the skew scattering contribution. The solid lines are electron propagators of band $+$ or $-$. The curved lines are phonon propagators. }
 \label{fig:extrinsic} 
\end{figure}

The anomalous charge Hall conductivity corresponds to the anti-symmetric part of Eq.(\ref{eq:Hall_conductivity})~\cite{Sinitsyn2007, Chen2023}. Since the total intrinsic charge AHE vanishes in monolayer TMD due to the TRS of the electron Hamiltonian, we  focus on the extrinsic contribution from chiral phonon scatterings, which includes a side-jump contribution and a skew scattering contribution~\cite{Sinitsyn2007, Chen2023}. Expanding the Feynman diagram in Fig.\ref{AH_conductivity}(a) in the electron eigenstate basis gives the Feynman diagrams of the side jump and skew scattering contributions as shown in Fig.\ref{fig:extrinsic}(a) and (b) respectively~\cite{Chen2023}, where the $+$ and $-$ lines correspond to the electron GF of the conduction and valence band respectively. Since the spin $z$ component is conserved, the interband scattering in each valley does not change the spin polarization and only one band of the conduction bands in Fig.\ref{fig:band} participates in the interband scatterings.  The leading order AHE contains one and only one conduction band propagator in each Feynman diagram as shown in Fig.\ref{fig:extrinsic} which corresponds to the interband scattering due to the e-phonon interaction and results in the AHE~\cite{Sinitsyn2007}.

The first four Feynman diagrams in Fig.\ref{fig:extrinsic}(a) give the same contribution to the side jump Hall conductivity. The Hall conductivity  corresponding to the first two diagrams from valley $\tau$ is 
\begin{eqnarray}\label{eq:sj-I}
\sigma_{xy}^{\rm sj, (2)}
&=&\int\frac{d^2 \mathbf{k}}{(2\pi)^3}\Big[
 (\delta\Gamma_x)^{-+}(\epsilon, \mathbf{k})G^R_{+}(\epsilon, \mathbf{k}) j_{y}^{+-}(\mathbf{k})G_{-}^A (\epsilon, \mathbf{k})\nonumber\\
 &&\quad +(\delta\Gamma_x)^{+-}(\epsilon, \mathbf{k})G_{-}^R(\epsilon, \mathbf{k}) j_{y}^{-+}( \mathbf{k})G^A_{+}(\epsilon, \mathbf{k})\Big]
\end{eqnarray}
where 
\begin{align}
 j_{y}^{+-}(\mathbf{k}) &\equiv  \langle +|\hat{j}_y|-\rangle= -\tilde{m} \sin \phi +i\tau \cos \phi, \\
 j_{y}^{-+}(\mathbf{k}) & \equiv  \langle -|\hat{j}_y|+ \rangle= -\tilde{m} \sin \phi -i\tau \cos \phi, \\
 G^{R/A}_{+}(\epsilon, \mathbf{k}) & \equiv  \langle +|\hat{G}^{R/A}_{\tau s}  |+\rangle =\frac{1}{\epsilon-\epsilon_{\tau s, +}(\mathbf{k})\pm i\gamma_+}, \\
G^{R/A}_{ -}(\epsilon, \mathbf{k}) &  \equiv  \langle -|\hat{G}^{R/A}_{\tau s}  |-\rangle =\frac{1}{\epsilon-\epsilon_{\tau s,-}(\mathbf{k})\pm i\gamma_-}, \\
 (\delta\Gamma_x)^{-+}(\epsilon, \mathbf{k}) & =  \sum_{r,\chi}\int\frac{d^2 \mathbf{q}}{(2\pi)^2}
\frac{F(\frac{\omega_{\mathbf{q}\chi}}{T}, \frac{\epsilon}{T})}{2\omega_\mathbf{q}}
 G^R_{-}(\epsilon', \mathbf{k'})G^A_{-}(\epsilon', \mathbf{k'})
  \nonumber\\
&\quad  \Gamma^{--}_x(\epsilon', \mathbf{k}')\ _\mathbf{k}\langle -|\mathcal S_{r,\chi,\tau,-}(\mathbf k')|+\rangle_\mathbf{k} 
\end{align}
with $\mathbf{k}=k(\cos\phi, \sin\phi), \mathbf{k'}=\mathbf{k}-\mathbf{q}\approx k(\cos\phi', \sin\phi')$, $ \phi'= \phi+\pi+2\theta,
\epsilon'=\epsilon_{\tau s,-}(\mathbf{k})+r\omega_{\mathbf{q},\chi}$  and $k=k_F, \epsilon=\epsilon_{\tau s,-}(\mathbf{k})=\epsilon_F$.
Using
\begin{eqnarray}
G^R_-(\epsilon', \mathbf{k'})G^A_-(\epsilon', \mathbf{k'})&\approx &\frac{\pi}{\gamma_-(\epsilon', \mathbf{k'})}\delta[\epsilon'-\epsilon_{\tau s,-}(\mathbf{k'})], \\
\ \ _\mathbf{k}\langle -|\mathcal S_{r\chi,\tau,-}(\mathbf k')|+\rangle_\mathbf{k}&\approx& g^2_d q^2 \frac{vk}{E(\mathbf{k})}\left\{\tilde{m} x^2+i \tau x \sqrt{1-x^2} \right. \nonumber\\
&& \left.-\rho^2[\tilde{m} + r(\chi\tau-\frac{\omega_{\mathbf{q},\chi}}{E_{\tau s}(\mathbf{k})})]\right\},
\end{eqnarray}
we get up to the first order of $\omega_{\mathbf{q,\chi}}$, 
\begin{eqnarray}\label{delta_Gamma_x}
&&(\delta\Gamma_x)^{-+}(\epsilon,\mathbf{k})=\frac{\kappa^2}{b_0-b_1}
 \sum_\chi\int_0^1dx\,
 \frac{x^2 F(\frac{\omega_{\mathbf{q}\chi}}{T}, \frac{\epsilon}{T})}{w(x) \sqrt{1-x^2}}\nonumber\\
&&
 \quad \quad \left\{\tilde{m} x^2(1-2x^2)\cos\phi
       +2i\tau x^2(1-x^2)\sin\phi \right.\nonumber\\
     &&\left. \quad \quad \quad -\rho^2\left[\tilde{m} +\tau \chi \frac{\omega_{\mathbf{q},\chi}}{E_{\tau s}(\mathbf{k})}\right](1-2x^2)\cos\phi\right\}
\end{eqnarray}
and ($\delta\Gamma_x)^{+-}=[(\delta\Gamma_x)^{-+}]^*$.

Combing the results of Eq.(\ref{eq:sj-I})-(\ref{delta_Gamma_x}), we get the side jump Hall conductivity in one valley due to the first four diagrams in Fig.\ref{fig:extrinsic}(a) as 
\begin{equation}\label{sj_I}
\frac{\sigma_{xy}^{\rm sj, I}}{e^2/h}
=-\frac{\tau\kappa^2}{2(b_0-b_1)}
[\tilde{m} I_1+\rho^2\tilde{m} (I_0-2I_1)+\rho^2 \tau I_3)],
\end{equation}
where 
\begin{eqnarray}\label{eq:I_0_1}
 I_0&=&\sum_\chi\int_0^1dx\,
 \frac{x^2 }{w(x)\sqrt{1-x^2}}F(\frac{\omega_{\mathbf{q},\chi}}{T}, \frac{\epsilon}{T})=\tilde{\xi}^d_0, \ \  \\
I_1&=&\sum_\chi\int_0^1dx\,
 \frac{x^4 }{w(x)\sqrt{1-x^2}}F(\frac{\omega_{\mathbf{q},\chi}}{T}, \frac{\epsilon}{T})=\frac{\tilde{\xi}^d_0-\tilde{\xi}^d_1}{2},\ \ \ \label{I_1} \\
I_3&=&\sum_\chi \chi \int_0^1dx\,
 \frac{x^2 (1-2 x^2) }{w(x)\sqrt{1-x^2}}\frac{\omega_{\mathbf{q}, \chi}}{E_{\tau s}(\mathbf{k})}F(\frac{\omega_{\mathbf{q}\chi}}{T}, \frac{\epsilon}{T}),\ \ \label{I_3}
\end{eqnarray}
and $b_0, b_1$ are given in Eq.(\ref{b_0})-(\ref{b_1}). For all the parameters in Eq.(\ref{sj_I})-(\ref{I_3}), $\epsilon=\epsilon_F, k=k_F, E_{\tau s}(\mathbf{k})=\sqrt{v^2 k_F^2+m^2_{\tau s}}$.

\begin{figure*}[htbp]
 \includegraphics[width=0.33\textwidth]{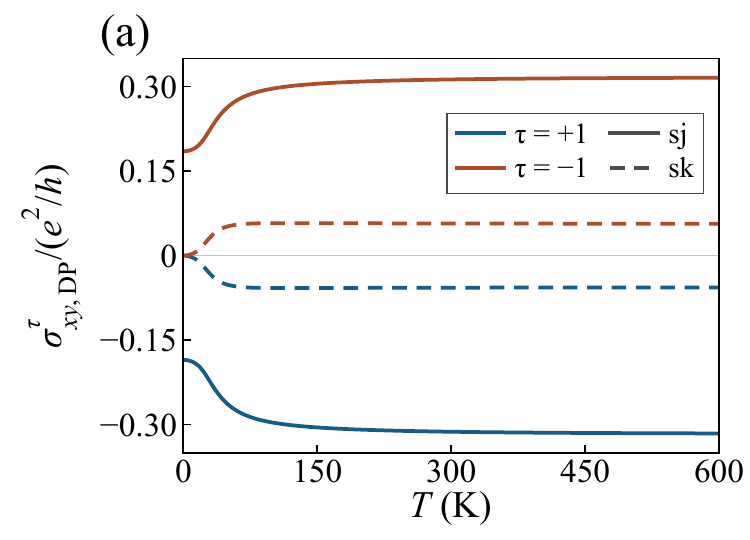} 
  \includegraphics[width=0.33\textwidth]{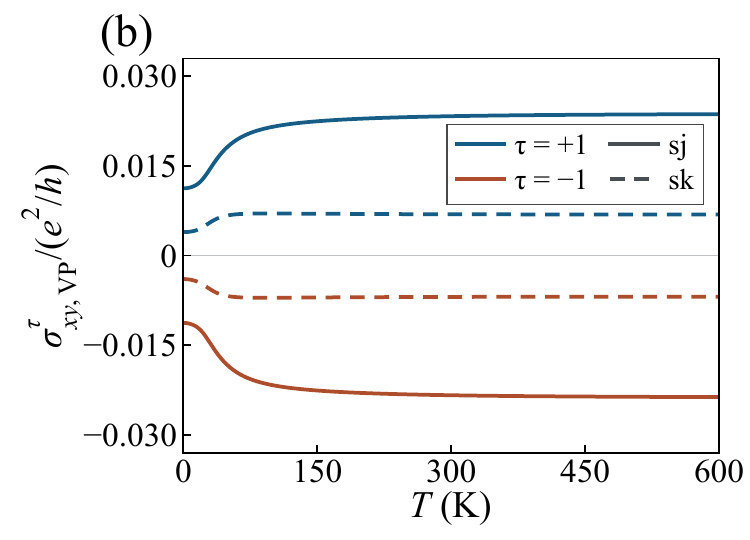} 
   \includegraphics[width=0.32\textwidth]{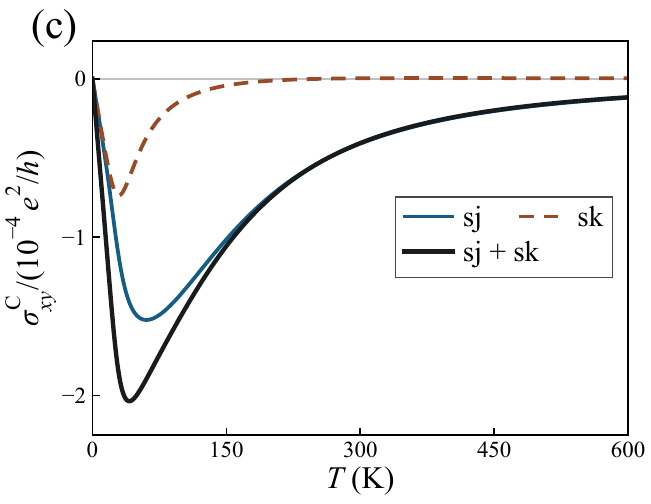} 
 \caption{\justifying(a)The anomalous Hall conductivity $\sigma^\tau_{xy}$ of valley $\tau$ from the DP scatterings as a function of temperature $T$. The solid and dashed lines represent the  side jump and skew scattering contribution respectively. The blue and brown lines represent the contribution from valley $\tau=+1$ and $\tau=-1$ respectively. (b)The anomalous Hall conductivity $\sigma^\tau_{xy}$ of valley $\tau$ from the VP scatterings as a function of temperature $T$. The labeling in the inset of (a) also applies to (b). (c)The total anomalous charge Hall conductivity $\sigma^C_{xy}$ of the two valleys as a function of temperature $T$. The blue solid line and brown dashed line represent the side jump and skew scattering contribution respectively. The black solid line presents the total anomalous charge Hall conductivity. The parameters of the electron and phonon band structure and interaction are the same as those in Fig.\ref{fig:Z}. }\label{fig:AH_conductivity}
\end{figure*}

The last four diagrams in Fig.\ref{fig:extrinsic}(a) give the same contribution to the side jump Hall conductivity. We then only need to compute one of them. The Hall conductivity due to the first of these four diagrams is 
\begin{equation}
\sigma_{xy}^{\rm sj,5}
=\int\frac{d^2 \mathbf{k}}{(2\pi)^3}
 \Gamma_x^{--}(\epsilon, \mathbf{k})(\delta G^R)_{-+}(\epsilon, \mathbf{k})j_{y}^{+-}(\mathbf{k})G_-^A(\epsilon, \mathbf{k}),
\label{eq:sj-e-kubo}
\end{equation}
where 
\begin{eqnarray}
 (\delta G^R)_{-+}(\epsilon, \mathbf k)
&=& G_-^{R}(\epsilon, \mathbf k)\,\Sigma^{R}_{-+}(\epsilon, \mathbf k)\,G_+^{R}(\epsilon, \mathbf k), \nonumber\\
 \Sigma^{R}_{-+}(\epsilon, \mathbf k)\equiv i \gamma_{-+}&= &\langle -|\Sigma_{\tau s}^{R}|+ \rangle=-i(\tilde{m} \Sigma_1-\kappa \Sigma_z). \nonumber
\end{eqnarray}
We then get the side jump Hall conductivity of this diagram as 
\begin{equation}\label{sj_5}
\frac{\sigma_{xy}^{\rm sj, 5}}{e^2/h}=- \frac{\tau \kappa \gamma_{-+}(\epsilon, \mathbf k)}{8 \gamma_0(b_0-b_1)}=-\frac{\tau \kappa^2}{8(b_0-b_1)}(\tilde{m} I_1-\rho^2 \tilde{\xi}^p_z).
\end{equation}
For all the parameters in Eq.(\ref{eq:sj-e-kubo})-(\ref{sj_5}), $(\epsilon, \mathbf{k})$ take the values on the Fermi surface, i.e, $\epsilon=\epsilon_F, k=k_F$.

The total contribution of the last four diagrams to the side jump Hall conductivity in Fig.
\ref{fig:extrinsic}(a) is then
\begin{equation}
\frac{\sigma_{xy}^{\rm sj, II}}{e^2/h}=-\frac{\tau \kappa^2}{2(b_0-b_1)}(\tilde{m} I_1-\rho^2 \tilde{\xi}^p_z).
\end{equation}

The side jump Hall conductivity of all the diagrams in Fig.\ref{fig:extrinsic}(a) from valley $\tau$ is then 
\begin{eqnarray}
&&\sigma_{xy}^{\rm sj, \tau}=\sigma_{xy}^{\rm sj, I}+\sigma_{xy}^{\rm sj, II}  \nonumber\\
&=&-\frac{\tau \kappa^2}{(b_0-b_1)}\left\{\tilde{m} I_1+\frac{\rho^2}{2}\left[\tilde{m} (I_0-2 I_1) -\tilde{\xi}^p_z +\tau I_3 \right] \right\}  \nonumber\\
&=& -\frac{\tau\kappa^2 \tilde{m}}{b_0-b_1}
 \bigg[I_1-\rho^2\sum_\chi\int_0^1 
 \frac{dx x^4 F(\frac{\omega_{\mathbf{q}\chi}}{T}, \frac{\epsilon}{T})}{w(x) \sqrt{1-x^2}}
 (1 +\tau\chi\frac{\omega_{\mathbf{q}, \chi}}{m_{\tau s}})\bigg], \label{sj_tau} \nonumber\\
\end{eqnarray}
where $\epsilon=\epsilon_F, k=k_F$ in all the parameters $\kappa, \tilde{m}, b_0, b_1$. The first term in the bracket comes from the DP scatterings and the second term comes from the VP scatterings.

The total side jump contribution from the two valleys is 
\begin{equation}\label{eq:sj_total}
\sigma_{xy}^{\rm sj}=\sum_{\tau=\pm 1}\sigma_{xy}^{\rm sj, \tau}.
\end{equation}

The side jump contribution to the anomalous charge Hall conductivity in each valley from the DP and VP scatterings is shown by solid lines in Fig.\ref{fig:AH_conductivity}(a) and (b) respectively. From Eq.(\ref{sj_tau}) and (\ref{I_1}), we can see that for DP scatterings, the anomalous Hall conductivity from left and right chiral phonons in one valley has the same sign. The  side jump Hall conductivity in a single valley due to DP scatterings is very close to that due to the linear-phonon-electron scatterings in 2D massive Dirac model in \cite{ZhangChen2024}. Specifically, at both $T\to 0$ and $T\to \infty$, the side jump Hall conductivity in one valley due to DP couplings of chiral phonon and electrons reduces to that of linear phonon-electron couplings in a 2D massive Dirac model in \cite{ZhangChen2024}. because in these two limits, the frequency splitting of left and right chiral phonons is negligible and a left and a right chiral phonon superpose to a linear phonon. Equation (\ref{sj_tau}) also shows that for purely DP couplings, the side jump Hall conductivity is opposite in the two valleys so the total side jump contribution to the charge Hall conductivity of the two valleys is zero.  

 In contrast, for VP coupling, part of the side jump Hall conductivity has the same sign in the two valleys and the sum in the two valleys is finite. For this part, the phonons with opposite chiralities lead to side jump Hall conductivity with opposite signs. Without frequency splitting of the left and right chiral phonons, the contributions from the phonons with opposite chirality cancel each other because the TRS is not broken. However, with the chiral phonon frequency splitting, the side jump Hall conductivity from left and right chiral phonons does not cancel each other. The total side jump Hall conductivity  is finite, as shown in Fig.\ref{fig:AH_conductivity}(c).

We note that since the vertex correction includes both DP and VP scatterings, the total side jump Hall conductivity in Eq.(\ref{eq:sj_total}) also includes processes of mixed DP and VP scattering though the Hall conductivity  with only DP scatterings vanishes.

We next compute the AH conductivity due to the skew scatterings. The Feynman diagrams of the skew scattering contribution are shown in Fig.\ref{fig:extrinsic}(b). The contribution from the first two diagrams is 
\begin{eqnarray}
\sigma_{xy}^{\mathrm{sk}, \rm I}
&=&\frac{1}{2\pi}\int\frac{d^2 \mathbf{k}}{(2\pi)^2}\Big[
 (\delta\Gamma_x)^{-+} G^R_+(\delta\Gamma_y)^{+-}G_-^A\notag\\
 &&\quad \quad \quad \quad +(\delta\Gamma_x)^{+-}G_-^R(\delta\Gamma_y)^{-+}G_+^A\Big]
\label{eq:hall-sk-kubo}
\end{eqnarray}
where $(\delta\Gamma_x)^{-+}, (\delta\Gamma_x)^{+-}$ and $G^{R/A}_{\pm}$ have been given in the previous text and  for brevity we have dropped the variables $(\epsilon, \mathbf{k})$ which take the values on the Fermi surface for these quantities. To get $(\delta\Gamma_y)^{-+}$, one only needs to exchange the $\sin\phi$ and $\cos\phi$ in $(\delta\Gamma_x)^{-+}$ in Eq.(\ref{delta_Gamma_x}), and $(\delta\Gamma_y)^{+-}=[(\delta\Gamma_y)^{-+}]^*$.  We get 
\begin{align}
&\sigma_{xy}^{\mathrm{sk},\rm I}=\frac{\tau\kappa^4 (I_1-I_2)}{(b_0-b_1)^2}
 \bigg[\tilde{m}(I_1-2I_2)
 \notag\\
&\quad -\rho^2\sum_\chi\int_0^1 dx\,
 \frac{x^2 (1-2x^2)F(\frac{\omega_{\mathbf{q}\chi}}{T}, \frac{\epsilon}{T})}{w(x)\sqrt{1-x^2}} 
 (\tilde{m}+\tau\chi\frac{\omega_{\mathbf{q},\chi}}{E_{\tau s}(\mathbf{k})})\bigg]
\end{align}
where $I_1$ is given in Eq.(\ref{I_1}) and 
\begin{equation}
I_2=\sum_\chi\int_0^1dx\,
 \frac{x^6  F(\frac{\omega_{\mathbf{q}\chi}}{T}, \frac{\epsilon}{T})}{w(x)\sqrt{1-x^2}}.
\end{equation}

The skew scattering contribution from the last four diagrams of Fig.\ref{fig:extrinsic}(b) 
is 
\begin{eqnarray}
\sigma_{xy}^{\mathrm{sk},\rm II}
&=&\frac{4}{2\pi}\int\frac{d^2 \mathbf{k}}{(2\pi)^2}
 \Gamma^{--}_x(\delta G^R)_{-+}(\delta\Gamma_y)_{+-}G_-^A \nonumber\\
 &=&-\frac{\tau\kappa^4 (I_1-I_2)}{(b_0-b_1)^2}
 (\tilde{m}I_1-\rho^2 \tilde{\xi^p_z}).
\end{eqnarray}

The total skew scattering  contribution of all the diagrams  in Fig.\ref{fig:extrinsic}(b) from valley $\tau$ is then
\begin{eqnarray}\label{sk_tau}
\sigma_{xy}^{\rm sk, \tau}&=&\sigma_{xy}^{\rm sk, I}+\sigma_{xy}^{\rm sk, II}=-\frac{2\tau\kappa^4(I_1-I_2)}{(b_0-b_1)^2}
 \bigg[\tilde{m} I_2 \nonumber\\
&& - \rho^2\sum_\chi\int_0^1 dx\,
 \frac{x^4 F(\frac{\omega_{\mathbf{q}\chi}}{T}, \epsilon)}{w(x)\sqrt{1-x^2}}
 (\tilde{m}+\tau\chi \frac{\omega_{\mathbf{q},\chi}}{E_{\tau s}(\mathbf{k})})\bigg]. \nonumber\\
\end{eqnarray}
The first term in the bracket comes from DP scattering and the second term from the VP scattering.
Similar to the side jump contribution, with only DP scattering, the skew scattering contribution in the two valleys is opposite and the sum of them vanishes. Whereas for VP, the skew scattering contribution in the two valleys is not completely opposite. For the reason, the total skew scattering contribution of the two valleys is finite. 
The skew scattering contribution to the anomalous charge Hall conductivity in each valley from the DP and  VP scatterings as a function of temperature $T$ are shown in Fig.\ref{fig:AH_conductivity}(a) and (b) by dashed lines respectively. 

The total anomalous Hall conductivity as a function of temperature is plotted in FIg.\ref{fig:AH_conductivity}(c).
At both $T\to 0$ and $T\to \infty$, the total anomalous charge Hall conductivity goes to zero because in both limits the frequency splitting of the left and right chiral phonons is negligible and the time reversal symmetry of the whole system is restored.

\subsection{The valley and spin Hall conductivity}

Other than the anomalous charge Hall effect, the system possesses significant valley Hall effect due to the inversion symmetry breaking~\cite{Xiao2007}. The valley Hall conductivity is 
\begin{equation}
\sigma^V_{xy}=\sum_\tau \tau \sigma^\tau_{xy},
\end{equation}
where $\sigma^\tau_{xy}=\sigma^{\rm int, \tau}_{xy}+ \sigma^{\rm sj, \tau}_{xy}+\sigma^{\rm sk, \tau}_{xy}$ is the  charge Hall conductivity in valley $\tau$ and includes both the intrinsic and extrinsic contribution, the latter includes the side jump and skew scattering  contribution in Eq.(\ref{sj_tau}) and (\ref{sk_tau}). 

\begin{figure}[t]
    \centering
    \includegraphics[width=0.45\textwidth]{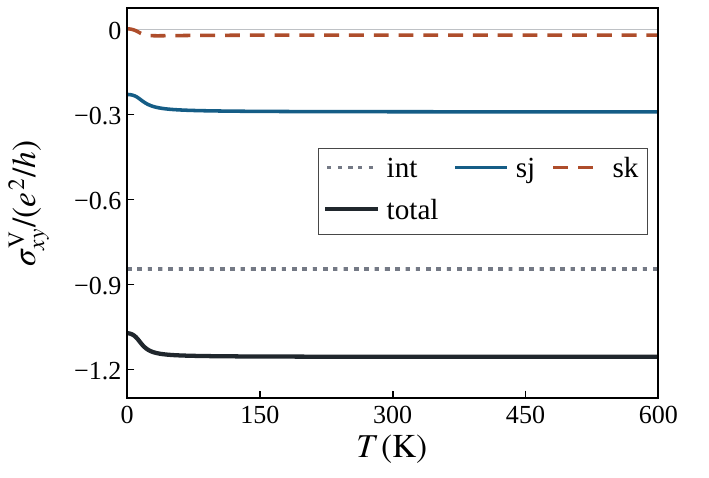}
    \caption{\justifying The valley Hall conductivity $\sigma^V_{xy}$ of the monolayer TMD as a function of temperature $T$. The dotted, dashed and blue solid lines represent the intrinsic, skew scattering and side jump contribution respectively. The solid black line represents the total valley Hall conductivity. The parameters of the electron and phonon band structure and interaction are the same as those in Fig.\ref{fig:Z} and Fig.\ref{fig:VH_conductivity}}
\label{fig:VH_conductivity}
\end{figure} 

The intrinsic Hall conductivity from valley $\tau$ is 
\begin{eqnarray}\label{eq:intrinsic}
\sigma_{xy}^{\rm int, \tau}
&=&\int\frac{d^2\mathbf{k}}{(2\pi)^3}[
 j_{x}^{-+}(\mathbf{k})G^{0,R}_+(\epsilon, \mathbf{k}) j_{y}^{+-}(\mathbf{k}) G_-^{0,A}(\epsilon, \mathbf{k}) \nonumber\\
&&
\quad +\ j_{x}^{+-}(\mathbf{k})G^{0,R}_-(\epsilon, \mathbf{k}) j_{y}^{-+}(\mathbf{k}) G_+^{0,A}(\epsilon, \mathbf{k})],
\end{eqnarray}
where $G^{0,R/A}_\pm$ is the bare electron GF of the $\pm$ band. We get 
\begin{equation}
\sigma_{xy}^{\rm int, \tau}=-\frac{\tau m_{\tau s}}{2 E_F}\frac{e^2}{h},
\end{equation}
which is opposite in the two valleys. 

The total intrinsic charge anomalous Hall conductivity is then zero, but the intrinsic valley Hall conductivity is finite as 
\begin{equation}
\sigma^{\rm V, int}_{xy}=-\frac{m_{\tau s}}{E_F}\frac{e^2}{h}.
\end{equation}

The side jump and skew scattering contributions to the valley Hall conductivity are respectively
\begin{eqnarray}
\sigma^{\rm V, sj}_{xy}&=&\sum_\tau \tau \sigma^{\rm sj, \tau}_{xy}, \\
\sigma^{\rm V, sk}_{xy}&=&\sum_\tau \tau \sigma^{\rm sk, \tau}_{xy}.
\end{eqnarray}

The temperature dependence of $\sigma^{\rm V, int}_{xy}, \sigma^{\rm V, sj}_{xy}, \sigma^{\rm V, sk}_{xy}$ and the total valley Hall conductivity $\sigma^{\rm V}_{xy}$ for given Fermi energy and chiral phonon dispersion is plotted in Fig.\ref{fig:VH_conductivity}.

Since the spin and valley is locked for Fermi energy locating at the valence band in Fig.\ref{fig:band}, the spin Hall conductivity 
\begin{equation}
\sigma^{\rm s}_{xy}=\sum_\tau s^\tau_z \sigma^\tau_{xy}=\frac12 \sigma^{\rm V}_{xy}.
\end{equation}
is also finite and equals to half of the valley Hall conductivity.

\section{Discussions and conclusion}\label{Discussion}
In the above calculation, we ignored the crossed-diagram skew scattering contribution~\cite{Titov2016, Chen2023_2}. For DP coupling of linear phonon and electrons, we have shown that the crossed-diagram  skew scattering is less important than the non-crossing skew scattering diagrams at low temperature $T$ for 2D massive Dirac model~\cite{ZhangChen2024}. Since at $T\to 0$, the chiral phonon-electron couplings reduces to the linear-phonon-electron couplings, we expect that the same is true for the DP coupling of chiral phonons and electrons. 
Moreover, since the DP is the same in the two valleys of monolayer TMD, we expect that the skew scattering contribution from the crossed diagrams in the two valleys also cancels each other. Yet the skew scattering contribution of the crossed diagrams due to VP couplings may still be finite and of the same order as the non-crossing diagrams. However, due to the complexity of crossed diagrams, we will leave the study of this contribution for a future work.

The chiral phonons in monolayer TMD include both acoustic and optical modes~\cite{ZhangNiu2015}. In this work, we only studied the interaction of electrons with acoustic phonon modes. The interaction between electrons and optical phonons may also results in electron anomalous Hall effect. However, the coupling of electrons and optical phonons is beyond the elastic theory and we will also leave the study of this topic for the future.

In conclusion, we studied the electron anomalous Hall effect in monolayer TMD induced by chiral phonon interaction with electrons. We show that though the electron Hamiltonian of this system preserves time reversal symmetry so the intrinsic anomalous Hall effect vanishes, the extrinsic contribution due to chiral phonon scatterings with electrons is finite. Specifically we studied two types of electron-phonon interactions, i.e., the DP coupling and VP coupling and showed that the DP induced anomalous charge Hall conductivity in the two valleys cancel each other whereas the VP induced anomalous charge Hall conductivity has different values in the two valleys so the total anomalous charge Hall effect is finite. The chiral phonon interaction with the electrons also results in significant valley and spin Hall effect in monolayer TMD due to the inversion symmetry breaking and spin-valley locking. Different from the charge Hall effect, the valley and spin Hall effects in the system include both intrinsic and extrinsic contribution.  


\section{Acknowledgment}
This work is supported by the National Natural Science Foundation of China under Grant No. 11974166 and the Natural Science Foundation of Jiangsu Province under Grant No.BK20231398.

\end{document}